\documentclass[paper,notoc]{JHEP3}
\usepackage{epsfig,cite}
\usepackage{amsbsy} 
\usepackage{epsfig}
\usepackage{graphicx}

\newcommand\new{\newcommand}         

\def\beq{\begin{equation}}   
\def\eeq{\end{equation}}
\def\bea{\begin{eqnarray}}  
\def\eea{\end{eqnarray}}

\new{\emem}{{\ifmmode\mathrm{e}^-\else e$^-$\fi}}
\new{\epem}{{\ifmmode\mathrm{e}^+\else e$^+$\fi}}
\new{\zbo}  {{\ifmmode\mathrm{Z}\else Z\fi}}
\new{\wpm} {{\ifmmode\mathrm{W}^\pm\else W$^\pm$\fi}}
\new{\wbo} {{\ifmmode\mathrm{W}\else W\fi}}
\new{\epm} {{\ifmmode\mathrm{e^+e^-}\else $\mathrm{e^+e^-}$\fi}}
\new{\qq}  {{\ifmmode\mathrm{q}\else q\fi}}
\new{\qqb} {{\ifmmode\bar{\mathrm{q}}\else $\bar{\mathrm{q}}$\fi}}
\new{\tq}  {{\ifmmode\mathrm{t}\else t\fi}}
\new{\tqb} {{\ifmmode\bar{\mathrm{t}}\else $\bar{\mathrm{t}}$\fi}}
\new{\bq}  {{\ifmmode\mathrm{b}\else b\fi}}
\new{\bqb} {{\ifmmode\bar{\mathrm{b}}\else $\bar{\mathrm{b}}$\fi}}
\new{\ttbar}{\tq\tqb}
\new{\qqbar}{\qq\qqb}
\new{\gu}  {{\ifmmode\mathrm{g}\else g\fi}}
\new{\qqbarg}{\qq\qqb\gu}
\new{\pp}  {{\ifmmode\mathrm{p}\else p\fi}}

\new{\hh}  {{\ifmmode h\else $h$\fi}}
\new{\HH}  {{\ifmmode \mathrm{H}\else $\mathrm{H}$\fi}}
\new{\fe}  {{\ifmmode f\else $f$\fi}}
\new{\lp}  {{\ifmmode \ell\else $\ell$\fi}}
\new{\XX}  {{\ifmmode X\else $X$\fi}}
\new{\Vp}  {{\ifmmode V\else $V$\fi}}
\new{\Kzs} {{\ifmmode\mathrm{K}_\mathrm{S}^0\else $\mathrm{K}_\mathrm{S}^0$\fi}}
\new{\Kzl} {{\ifmmode\mathrm{K}_\mathrm{L}^0\else $\mathrm{K}_\mathrm{L}^0$\fi}}
\new{\Kp} {{\ifmmode\mathrm{K}\else $\mathrm{K}$\fi}}
\new{\ppHWW} {{\ifmmode\pp\pp\rightarrow\HH\rightarrow\wbo\wbo
                             \else $\pp\pp\rightarrow\HH\rightarrow\wbo\wbo$\fi}}
\new{\ppHWWlept} {{\ifmmode\pp\pp\rightarrow\HH\rightarrow\wbo\wbo\rightarrow\lp\nu\lp\nu
                             \else $\pp\pp\rightarrow\HH\rightarrow\wbo\wbo\rightarrow\lp\nu\lp\nu$\fi}}
\new{\HWWlept} {{\ifmmode\HH\rightarrow\wbo\wbo\rightarrow\lp\nu\lp\nu
                             \else $\HH\rightarrow\wbo\wbo\rightarrow\lp\nu\lp\nu$\fi}}

\new{\LEP}        {\mbox{\small\textsc{LEP}}}
\new{\LEPONE}     {\mbox{\small\textsc{LEP1}}}
\new{\LEPTWO}     {\mbox{\small\textsc{LEP2}}}
\new{\CERN}       {\mbox{\small\textsc{CERN}}}
\new{\ALEPH}      {\mbox{\small\textsc{ALEPH}}}
\new{\DELPHI}     {\mbox{\small\textsc{DELPHI}}}
\new{\LD}         {\mbox{\small\textsc{L3}}}
\new{\OPAL}       {\mbox{\small\textsc{OPAL}}}
\new{\SPS}        {\mbox{\small\textsc{SPS}}}
\new{\TEVATRON}   {\mbox{\small\textsc{TEVATRON}}}
\new{\LHC}        {\mbox{\small\textsc{LHC}}}
\new{\FERMILAB}   {\mbox{\small\textsc{FERMILAB}}}
\new{\CDF}        {\mbox{\small\textsc{CDF}}}
\new{\DZERO}      {\mbox{\small\textsc{D0}}}
\new{\CTEQ}        {\mbox{\small\textsc{CTEQ}}}
\new{\FNAL}        {\mbox{\small\textsc{FNAL}}}
\new{\ATLAS}        {\mbox{\small\textsc{ATLAS}}}
\new{\CMS}        {\mbox{\small\textsc{CMS}}}

\new{\eV}         {{\ifmmode {\mathrm{ eV}}\else ${\mathrm{ eV}}$\fi}}
\new{\MeV}        {{\ifmmode {\mathrm{ MeV}}\else ${\mathrm{ MeV}}$\fi}}
\new{\MeVc}       {{\ifmmode {\mathrm{ MeV}}/c\else ${\mathrm{ MeV}}/c$\fi}}
\new{\MeVcc}      {{\ifmmode {\mathrm{ MeV}}/c^2\else ${\mathrm{ MeV}}/c^2$\fi}}
\new{\GeV}        {{\ifmmode {\mathrm{ GeV}}\else ${\mathrm{ GeV}}$\fi}}
\new{\GeVc}       {{\ifmmode {\mathrm{ GeV}}/c\else ${\mathrm{GeV}}/c$\fi}}
\new{\GeVcc}      {{\ifmmode {\mathrm{ GeV}}/c^2\else ${\mathrm{GeV}}/c^2$\fi}}
\new{\TeV}        {{\ifmmode {\mathrm{ TeV}}\else ${\mathrm{ TeV}}$\fi}}

\new{\fb}        {{\ifmmode {\mathrm{ fb}}\else ${\mathrm{ fb}}$\fi}}
\new{\fbinv}   {{\ifmmode {\mathrm{ fb}^{-1}}\else ${\mathrm{ fb}^{-1}}$\fi}}
\new{\pb}        {{\ifmmode {\mathrm{ pb}}\else ${\mathrm{ pb}}$\fi}}
\new{\pbinv}   {{\ifmmode {\mathrm{ pb}^{-1}}\else ${\mathrm{ pb}^{-1}}$\fi}}

\new{\JS}         {\mbox{\small\textsc{JETSET}}}
\new{\HW}         {\mbox{\small\textsc{HERWIG}}}
\new{\AR}         {\mbox{\small\textsc{ARIADNE}}}
\new{\PY}         {\mbox{\small\textsc{PYTHIA}}}
\new{\JSv}        {\mbox{\small\textsc{JETSET\ 7.405}}}
\new{\HWo}        {\mbox{\small\textsc{HERWIG\ 5.8}}}
\new{\HWn}        {\mbox{\small\textsc{HERWIG\ 5.9}}}
\new{\ARv}        {\mbox{\small\textsc{ARIADNE\ 4.05}}}
\new{\PYv}        {\mbox{\small\textsc{PYTHIA\ 5.7}}}
\new{\FEHiP}        {\mbox{\small\textsc{FEHiP}}}
\new{\LFEHiP}        {\mbox{\textsc{FEHiP}}}
\new{\pvegas}        {\mbox{\small\textsc{PVEGAS}}}

\new{\Mz}         {{\ifmmode M_{\mathrm{ Z}}
                    \else $M_{\mathrm{ Z}}$\fi}}
\new{\Mzsq}       {{\ifmmode M^2_{\mathrm{ Z}}
                    \else $M^2_{\mathrm{ Z}}$\fi}}
\new{\Mw}         {{\ifmmode M_{\mathrm{ W}}
                    \else $M_{\mathrm{ W}}$\fi}}
                    
\new{\MH}         {{\ifmmode M_{\mathrm{ h}}
                    \else $M_{\mathrm{ h}}$\fi}}

\new{\as}[1]      {{\ifmmode\alpha^{#1}_s
                    \else$\alpha^{#1}_s$\fi}}
\new{\asx}[1]      {{\ifmmode a^{#1}_s
                    \else $a^{#1}_s$\fi}}
\new{\asb}[1]     {{\ifmmode\overline{\alpha}^{#1}_s
                    \else $\overline{\alpha}^{#1}_s$\fi}}
\new{\asmz}       {{\ifmmode\alpha_s(\Mzsq)
                    \else $\alpha_s(\Mzsq)$\fi}}
\new{\lqcd}       {{\ifmmode\Lambda_{\mathrm{ QCD}}
                    \else $\Lambda_{\mathrm{ QCD}}$\fi}}
\new{\lqcdsq}     {{\ifmmode\Lambda^2_{\mathrm{ QCD}}
                    \else $\Lambda^2_{\mathrm{ QCD}}$\fi}}
\new{\llla}       {{\ifmmode\Lambda_{\mathrm{ LLA}}
                    \else $\Lambda_{\mathrm{ LLA}}$\fi}} 
\new{\lmsbar}[1]  {{\ifmmode \Lambda^{(#1)}_{\overline{\mathrm{MS}}}
                    \else $\Lambda^{(#1)}_{\overline{\mathrm{MS}}}$\fi}}
\new{\lmsb}       {{\ifmmode \Lambda_{\overline{\mathrm{MS}}}
                    \else $\Lambda_{\overline{\mathrm{MS}}}$\fi}}
\new{\lmsbsq}     {{\ifmmode \Lambda^{2}_{\overline{\mathrm{MS}}}
                    \else $\Lambda^{2}_{\overline{\mathrm{MS}}}$\fi}}
\new{\pt}       {{\ifmmode p_{\mathrm{T}}
                    \else $p_{\mathrm{T}}$\fi}}
\new{\Etmiss}       {{\ifmmode E_{\mathrm{T}}^{\mathrm{miss}}
                    \else $E_{\mathrm{T}}^{\mathrm{miss}}$\fi}}
\new{\ptlmin}       {{\ifmmode p_{\mathrm{T}}^{\lp\mathrm{min}}
                    \else $p_{\mathrm{T}}^{\lp\mathrm{min}}$\fi}}
\new{\ptlmax}       {{\ifmmode \mathrm{p}_{\mathrm{T,max}}^{\mathrm{cut}}
                    \else $\mathrm{p}_{\mathrm{T,max}}^{\mathrm{cut}}$\fi}} 
\new{\ptlep}       {{\ifmmode p_{\mathrm{T}}^{\mathrm{lepton}}
                    \else $p_{\mathrm{T}}^{\mathrm{lepton}}$\fi}}  
\new{\ptjet}       {{\ifmmode p_{\mathrm{T}}^{\mathrm{jet}}
                    \else $p_{\mathrm{T}}^{\mathrm{jet}}$\fi}}                                      
\new{\ptveto}       {{\ifmmode p_{\mathrm{T}}^{\mathrm{veto}}
                    \else $p_{\mathrm{T}}^{\mathrm{veto}}$\fi}}          

\new{\phicut}       {{\ifmmode \phi_{\lp \lp}^{\mathrm{cut}}
                    \else $\phi_{\lp \lp}^{\mathrm{cut}}$\fi}}          

\new{\Etcut}       {{\ifmmode E_{\mathrm{T},\mathrm{miss}}^{\mathrm{cut}}
                    \else $E_{\mathrm{T},\mathrm{miss}}^{\mathrm{cut}}$\fi}}

\title{\boldmath NNLO QCD predictions for the \HWWlept\ signal at the LHC}

\author{Charalampos Anastasiou\\
PH-Department, CERN\\
1211 Geneva, Switzerland\\
	E-mail: \email{babis@cern.ch}}
\author{G\"unther Dissertori\\
Institute for Particle Physics, ETH Zurich,\\
       8093 Zurich, Switzerland\\
	E-mail: \email{dissertori@phys.ethz.ch}}
\author{Fabian St\"ockli\\
Institute for Particle Physics, ETH Zurich,\\
       8093 Zurich, Switzerland\\
	E-mail: \email{fabstoec@phys.ethz.ch}}

\abstract{ 
We present a first computation of the next-to-next-to-leading order (NNLO) QCD  
cross-section at the LHC for the production of four leptons 
from a Higgs boson decaying into $\wbo$ bosons. We study the 
cross-section for a mass value of $\MH = 165\,\GeV$; 
around this value a Standard Model Higgs boson decays 
almost exclusively into \wbo-pairs. We 
apply all nominal experimental cuts on the final state leptons and the 
associated jet activity and study the magnitude 
of higher order effects up to NNLO on all kinematic variables
which are constrained by experimental cuts.
We find that the magnitude of the higher order corrections 
varies significantly with the signal selection cuts. 
As a main result we give
 the value of the cross-section at NNLO with all selection cuts 
envisaged  for the search of the Higgs boson.
} 
\preprint{{ETHZ-IPP/PR-2007-01}, {CERN-PH-TH/2007-118}}

\begin{document}

\section{Introduction}
\label{sec:intro}

The search for the Higgs boson will be  one of the major 
experimental activities at the Large Hadron Collider. 
The \ATLAS\ and \CMS\ detectors
at the \LHC\ are designed to discover a Higgs boson with a mass up to about  
$1 \, \TeV$. 
The experimental signals of a Higgs boson
have been studied in detail during the last years.
These studies  indicate that a $5\,\sigma$  discovery of a Standard Model (SM) 
Higgs boson could be possible over  the entire mass range with an 
integrated luminosity of about $30\,\fbinv$ (see, for example, 
\cite{CMStdr}).

In the mass regions below $\sim 155\,\GeV$ and above $\sim 180\,\GeV$ the 
main detection channels  are $\HH\rightarrow\gamma\gamma$ and 
 $\HH\rightarrow\zbo\zbo\rightarrow 4\lp$, where narrow invariant mass 
peaks can be reconstructed from isolated photons and leptons. 
In the region between $155\,\GeV$ and  $180\,\GeV$ 
the Higgs boson decays almost exclusively into a pair of nearly 
on-shell $\wbo$ bosons, which subsequently decay to jets or 
lepton-neutrino pairs. 

The discovery of a Higgs boson in this mass range was for a long time regarded 
as very difficult. The hadronic and semi-leptonic channels are not viable for the 
discovery because of the overwhelming QCD jet background. 
The leptonic channel with two isolated charged leptons and large missing transverse 
energy provides a much cleaner signal, however, 
because of the undetected neutrinos in the 
final state no narrow mass peak can be reconstructed. The absence of the latter 
could be compensated by the large 
cross-section~\cite{Spira:1995rr,Dawson:1990zj,Harlander:2002wh,Anastasiou:2002yz,Ravindran:2003um} 
if the dominant backgrounds of non-resonant $\pp\pp \to \wbo \wbo$\ and  
$\pp\pp \to \ttbar$ production were reduced significantly. 
Before any selection cuts are applied, 
the top-quark background cross-section is about 45 times and 
the \wbo-pair background cross-section  about 6 times larger than the 
signal cross-section~\cite{DavatzCMSnote}. 
Good selection  criteria to reduce  these 
backgrounds were not found easily; it was believed for some 
time that a Higgs boson with a mass in this range could 
remain undetected at the LHC. 

In 1996, Dittmar and Dreiner \cite{Dittmar:1996ss} studied the effects 
of spin correlations  and the mass of the resonant and non-resonant 
\wbo\wbo\ system. For signal events they observed that the 
opening angle $\phi_{\lp\lp}$ 
between the leptons in the plane transverse 
to the beam axis tends to be small; in addition, the transverse 
momentum (\pt) spectrum of the charged leptons is 
somewhat sensitive to the Higgs-boson mass. 
In contrast, the  lepton angle $\phi_{\lp\lp}$ for the background tends to 
be large and can be used as a discriminating variable. 
In order to reduce the large top-pair background, which is 
characterized by strong jet activity, they  proposed to reject events 
where jets have a  large $\pt$.  
With these basic selection criteria,
it has been concluded that a discovery in the channel \HWWlept\ with 
$\lp = \mathrm{e},\mu,\tau\,(\rightarrow \lp\nu\nu)$ 
for a Higgs mass from $155\,\GeV$ to $180\,\GeV$
is indeed possible~\cite{Dittmar:1996ss}, even with only a few \fbinv\ of integrated 
luminosity~\cite{DavatzCMSnote}.

The ratio of the Higgs signal 
cross-section to the cross-section for the background processes 
after the application of such cuts is estimated to range between $1:1$ and $2:1$, 
depending on the precise value of the Higgs  boson mass. 
The tuning of the selection cuts which leads to 
these spectacular ratios~\cite{Dittmar:1996ss,DavatzCMSnote} 
is based on a thorough analysis of many kinematic distributions 
for both signal and background processes. The required 
cross-sections were 
calculated~\cite{Davatz:2006ja,Davatz:2006kb,Davatz:2006ka} using a 
leading-order parton 
shower Monte-Carlo simulation combined with 
re-weighting methods, in an attempt to effectively incorporate 
the effects of higher order QCD corrections~\cite{Davatz:2006ut,Davatz:2004zg}.

A precise knowledge of the  cross-sections and the efficiency of 
the selection cuts is particularly important in this discovery channel 
because of two reasons: 

(i) The cuts reduce the cross-section for the signal by one order 
of magnitude  and the 
background  by almost three orders of magnitude; 
a small uncertainty in the efficiency could result in a 
more significant uncertainty in the signal 
to background ($S/B$) ratio.   

(ii) Unlike other mass regions where a resonance mass peak 
can be reconstructed, the measurement of the Higgs boson mass will rely on 
the precise knowledge of both the signal cross-section and distributions of 
kinematic observables~\cite{Davatz:2006ja}.

The inclusive cross-section for the production of a Higgs boson at the 
LHC receives large corrections at next-to-leading-order 
(NLO)~\cite{Spira:1995rr,Dawson:1990zj}  and smaller but significant 
corrections at next-to-next-to-leading-order 
(NNLO)~\cite{Harlander:2002wh,Anastasiou:2002yz,Ravindran:2003um} in QCD. 
It is believed that  corrections beyond NNLO are small, 
as indicated by recently computed leading logarithmic 
contributions at NNNLO~\cite{Moch:2005ky,Ravindran:2006cg} and 
resummation~\cite{Bozzi:2007pn,Catani:2003zt, Kulesza:2003wn, 
Ravindran:2006bu}.

The computation of differential cross-sections beyond NLO 
is challenging. The first NNLO differential distribution 
for a collider process was computed in 
2003~\cite{Anastasiou:2003yy,Anastasiou:2003ds}.  Fully differential 
cross-sections have appeared soon after and a significant number of 
new results has been  published~\cite{Ridder:2007bj,Catani:2007vq,Melnikov:2006kv,Melnikov:2006di,Anastasiou:2005pn,Weinzierl:2006yt,Anastasiou:2004qd}.  
The cross-section for the production of a Higgs boson via gluon fusion 
$\pp\pp \to \HH$ was the first example of such a calculation for 
a hadron collider process~\cite{Anastasiou:2004xq}. 
An application of this result was the NNLO prediction for the 
di-photon Higgs signal cross-section at the 
LHC~\cite{fehip}. Recently,  a Monte-Carlo program for the same purpose, 
based on a different method for computing  NNLO cross-sections,  
has been presented in~\cite{Catani:2007vq}.

Comparisons of the NNLO results with those of the event 
generators PYTHIA and MC@NLO~\cite{pythia,herwig,mcnlo} 
for the di-photon signal~\cite{Stockli:2005hz,Davatz:2006ut}  
showed that, in most cases, higher order effects can be well 
approximated by multiplying  the predictions of the 
generators with the $K$-factor for the inclusive cross-section. 
However, the cuts for the di-photon signal are mild  and do not
alter significantly the shape of kinematic distributions, while the 
reduction of the Higgs boson cross-section by selection cuts 
like the ones discussed above is 
drastic in the \ppHWWlept\ channel. 
The distributions of kinematic observables after selection cuts 
may have very different properties than the corresponding inclusive
distributions.
An example for this behavior can be found in the study of the 
jet-veto at NNLO~\cite{Catani:2001cr,fehip}. Additional evidence 
is shown by re-weighting leading-order Monte-Carlo generator events 
with $K$-factors to account for higher order effects in kinematic 
distributions of the Higgs boson~\cite{Davatz:2006ut,Davatz:2004zg}.  
From these observations it becomes clear that it is essential to compute 
kinematic distributions of the final-state leptons and the signal 
cross-section with all experimental cuts applied at NNLO in QCD.

In Ref.~\cite{fehip}, the NNLO Monte-Carlo program \FEHiP\ was published. 
\FEHiP\ computes differential 
cross-sections for  Higgs boson production via 
gluon fusion and includes a selection 
function for applying experimental cuts on the di-photon final 
state.  
In this paper we 
extend \FEHiP\ to include the matrix-elements for the decay of the 
Higgs boson in the \ppHWWlept\ channel and a selection function for the 
leptonic final-state.  In addition, we have parallelized the evaluation of 
distinct contributions to the cross-section.  
The results of our paper comprise kinematic distributions of the final state 
leptons as well as the cross-section for \ppHWWlept\ at next-to-next-leading 
order of perturbative QCD, taking into account all selection cuts at parton level.

\section{The NNLO Monte-Carlo program \LFEHiP}
\label{sec:theory}

\FEHiP\ computes phase-space integrals with 
arbitrary selection cuts and infrared divergences due to unresolved 
single or double real radiation~\cite{Anastasiou:2004qd}. 
The NNLO matrix-elements for Higgs boson production in gluon fusion
are rendered numerically integrable, 
by applying a sector decomposition algorithm~\cite{Binoth:2000ps,Gehrmann-DeRidder:2003bm,Anastasiou:2004qd}, 
splitting the phase-space into sectors
with a simplified infrared structure. 

In this paper, 
we extend \FEHiP\ to the \ppHWWlept\ decay channel. This requires the 
decay matrix-elements for $\HH \to  \wbo\wbo  \to \lp\nu \lp\nu$
and a selection function for the four leptons  in the final state. 

There are two methods to combine the various sectors 
into the final result:
 
(i) We can add up the integrands for all sectors
before performing a Monte-Carlo integration; 
this has the advantage that large cancellations among sectors do not 
spoil the accuracy of the numerical integration. The drawback of this 
approach is that each sector exhibits a different 
singularity structure; the adaptation of the integration to the peaks of 
the combined integrand is then complicated.

(ii) We can integrate each sector independently and add up the 
results at the end. The integrands for each sector are now simpler, 
but large cancellations between positively and negatively valued 
sectors may spoil the statistical accuracy of the final result.

In Ref.~\cite{fehip} it was found that adding  the sectors together 
before  integration resulted in a better performance 
for a single (not decaying) Higgs boson or the photon pair  as  final states. 
In a non-parallel  computation (which was sufficient), 
the alternative to integrate the sectors separately was slow. 

In our  current calculation, the experimental cuts reject a large 
part of the total cross-section, and a very good sampling of the 
phase-space is required. 
This is prohibitively slow for  the sum of the sectors.  
We have modified \FEHiP\ in order to integrate each sector separately. 
We have found that 
the Monte-Carlo adaptation in each sector is excellent. We  
did not encounter large cancellations 
among sectors; the cross-sections for individual sectors 
were usually of the same order of magnitude as the final result.

We have performed a two-fold parallelization of \FEHiP. First, 
each sector is integrated on a dedicated set of independent  processor
units. Second, each sector may be integrated in parallel on up to 64 CPUs 
using a program based on the algorithm~\pvegas~\cite{pvegas}.
The parallelization of sector decomposition for 
the computation in this paper serves as a 
successful prototype example for other future applications 
of the method.

\section{Selection cuts and physical parameters}
\label{sec:selection}

In the following we describe the experimental cuts which we use in our studies. 
These cuts are required to isolate the Higgs signal from the background, 
as discussed in the introduction. 
We keep the values of the cut parameters as close as possible to 
the ones described in Refs~\cite{DavatzCMSnote,Davatz:2006ja} and in 
the \CMS\ Physics Technical Design Report~\cite{CMStdr}. 
These cuts are motivated by the original study 
of~\cite{Dittmar:1996ss}, but are not identical.
 
As a first selection two isolated leptons (electrons or muons) with opposite 
charge and high transverse momentum $\pt$ are required.
Such leptons mainly originate from decays of electro-weak gauge bosons.
In order to reject Drell-Yan \zbo-production events,
these leptons should not be back-to-back in the plane transverse to 
the beam axis and their invariant mass should be well below the \zbo\ mass. 
Furthermore, some missing transverse energy is required. 
After applying these selection criteria the remaining sample is dominated by 
events which contain a pair of charged leptons originating from the decay of 
\wbo s, either from the signal or from the main backgrounds. 
The parameters we consider for this first selection ({\it pre-selection cuts}) are:
\begin{enumerate}
\item both charged leptons should have a transverse momentum of 
$\pt > 20\,\GeV$ and a pseudorapidity $|\eta| < 2$;
\item the di-lepton mass should be  $M_{\lp\lp} < 80\,\GeV$; 
\item the missing energy in the event, \Etmiss, 
has to exceed $20\,\GeV$\footnote{We compute \Etmiss ~from the momenta of the 
neutrinos. In a real experiment this variable must be computed differently. 
One possibility is to compute it by balancing the \pt~of the visible leptons. 
This is a relatively 
accurate approach when a jet-veto is applied, since 
it forbids any large jet activity in the central region. 
We have observed that defining $\Etmiss$ from the momenta of the 
neutrinos or the momenta of the visible leptons 
yields results which differ by less than $3\%$ at NLO when 
all other cuts for signal selection are applied.}; 
\item the opening angle $\phi_{\lp\lp}$ between the two leptons in the 
transverse plane should be smaller than $135^{\circ}$. 
\end{enumerate}

Following this pre-selection, further kinematic cuts exploit the 
different dynamics in signal and background~: 
(i) \wbo-pairs
from top-quark decays  are usually 
accompanied by jets, therefore a jet-veto can strongly reduce 
the \tq\tqb\ background; 
(ii) spin correlations lead to a small opening angle for signal events, 
in contrast to the non-resonant \wbo-pair production, and 
(iii) for the signal the observable lepton transverse momentum spectra show a 
Jacobian peak-like structure which depends on the Higgs mass.

We consider the following more stringent experimental
cuts, which are designed to isolate the Higgs signal 
({\it signal cuts}): 
\begin{enumerate}
\item the charged leptons should have a transverse momentum of 
$\pt > 25\,\GeV$ and a pseudorapidity $|\eta|<2$;
\item these leptons must be isolated from hadrons; 
the hadronic energy within a cone of $R=0.4$ around each lepton must not 
exceed $10\%$ of the corresponding lepton transverse momentum; 
\item the di-lepton mass should fall into the 
range $12\,\GeV < M_{\lp\lp} < 40\,\GeV$. 
The lower cut reduces potential backgrounds from \bq-resonances;
\item the missing transverse energy in the event, \Etmiss, 
has to exceed $50\,\GeV$; 
\item the opening angle $\phi_{\lp\lp}$ between the two leptons in the 
transverse plane should be smaller than $45^{\circ}$; 
\item there should be no jet with a transverse momentum larger than 
$25\,\GeV$~\footnote{In~\cite{DavatzCMSnote} a cut on the un-corrected transverse energy and a jet sub-structure parameter are used which corresponds to a jet transverse-energy cut of about $25\,\GeV$.} 
and  pseudorapidity   $|\eta|< 2.5$. 
Jets are found using a cone algorithm with a cone size of $R = 0.4$;
\item the harder lepton is required to have $30\,\GeV < p_\mathrm{T}^{\mathrm{lept}} < 55\,\GeV$.
\end{enumerate}

In what follows we study a Higgs boson with a mass value 
$M_h= 165 \, \GeV$; the width of the Higgs boson is computed to 
be $0.254 \, \GeV$ using the program HDECAY~\cite{hdecay}. The Higgs 
propagator is treated in the narrow width approximation. By comparing with
MCFM~\cite{mcfm}, which includes a Breit-Wigner distribution for the 
Higgs propagator, we found that 
at LO and NLO this is accurate within $2\%$. We have set 
$M_\mathrm{W} = 80.41 \, \GeV$ and take into account finite width effects for 
the \wbo\ bosons; we set $\Gamma_\mathrm{W} =2.06 \, \GeV$. 
The mass of the top-quark is set to $M_\mathrm{t}t = 175  \, \GeV$. \FEHiP\ 
calculates the Higgs boson cross-section in the infinite top-quark mass 
approximation, but the result is normalized to the  Born 
cross-section with the exact top-quark mass dependence 
(the b-quark contribution to the Born amplitude is neglected). 
We are using the MRST2001~\cite{mrst2001} at LO 
and the MRST2004~\cite{mrst2004} parton distribution functions 
at NLO and NNLO. 

All cross-sections which we present in the rest of the 
paper, correspond to one final-state 
lepton combination, e.g. 
$\pp\pp \to \HH +X \to \wbo^{+} \wbo^{-} + X\to \mathrm{e}^{+} \mathrm{e}^{-} \nu \bar{\nu} + X$.
In order to obtain the cross-sections for combinations of 
lepton final-states our results  
need to be multiplied with a factor $4$ for 
all $(\mathrm{e}, \mu)$  combinations and with a factor $9$ for all  
$(\mathrm{e},\mu, \tau)$ combinations~\footnote{We do  not consider the decay of the $\tau$ leptons.}.

In this work we only study the production of a Higgs boson 
in gluon fusion, without considering the weak boson fusion 
process~\cite{Rainwater:1999sd,Figy:2003nv}. 
We also do not consider  the effect of 
electroweak corrections to the production~\cite{Degrassi:2004mx} 
or the decay of the Higgs boson~\cite{Bredenstein:2006ha}.
The process $\pp\pp \to \mathrm{Z}\mathrm{Z} \to \lp \nu \lp \nu$ and interference 
effects will be the subject of a future publication.

In Section~\ref{sec:LHCresults} we shall present the 
cross-section for both the {\it pre-selection cuts}  
and the {\it signal cuts}.

\section{Magnitude of QCD corrections for kinematic 
distributions}
\label{sec:distributions}

In this Section we study the cross-section through NNLO, 
applying a cut on only one kinematic variable at a time.
In all plots of this Section, we consider a typical variation of 
the renormalization 
($\mu_\mathrm{R}$) and factorization scale  ($\mu_\mathrm{F}$) 
simultaneously, in the range 
$\frac{\MH}{2} < \mu = \mu_\mathrm{R} = \mu_\mathrm{F} < 2 \MH$. 
The inclusive cross-section for $\pp\pp \to \HH + X \to \lp \nu \lp \nu + X$  is 
given in Table~\ref{tab:inclusive}.
\begin{table}[h]
\begin{center}
\begin{tabular}{||c|c|c|c||}
\hline
$\sigma(\mathrm{fb})$         & LO     & NLO    &   NNLO  \\ 
\hline
$ \mu=\frac{\MH}{2}$  & $152.63 \pm 0.06$  
                   & $270.61 \pm 0.25$
                   & $301.23 \pm 1.19$ 
                   \\
\hline
$\mu= 2 \MH$       &    $103.89 \pm 0.04$  
                   & $199.76 \pm 0.17$ 
                   & $255.06 \pm 0.81$
                   \\
\hline
\end{tabular}
\end{center}
\caption{
\label{tab:inclusive}
The cross-section through NNLO with no experimental cuts applied.}
\end{table}
The $K$-factors for the inclusive cross-section, 
\begin{equation}
K_{\mathrm{(N)NLO}}(\mu) = \frac{\sigma_{\mathrm{(N)NLO}}(\mu)}{\sigma_{\mathrm{LO}}(\mu)}, 
\label{eq:Kfactor}
\end{equation}
range from 1.77 to 1.92 at NLO and from 1.97 to 2.45 at NNLO, 
depending on the scale choice~\footnote{
Note that the $K$-factor is often defined in the literature 
as the ratio of the NLO or the  
NNLO cross-section at a scale $\mu$ over the LO cross-section 
at a fixed scale $\mu_0$ (e.g. $\mu_0 = M_h$). 
Since we allow with our definition in 
Eq.~\ref{eq:Kfactor} both numerator and denominator to vary, 
a large scale variation of the 
$K$-factor does not necessarily indicate a big scale variation of the  
NLO or the NNLO cross-section in the numerator.}. 

It is important to compare the perturbative expansions for the 
inclusive cross-section and differential 
Higgs boson observables. We find many kinematic distributions 
which exhibit a different perturbative pattern than the inclusive 
cross-section. We present here integrated differential distributions 
$$
\sigma(X) = \int^X \frac{\partial \sigma}{ \partial x} dx; 
$$
the result for a bin $x \in [X_1, X_2]$ can be obtained from the 
difference 
$$
\sigma(x \in [X_1,X_2] )  = \sigma(X_2) - \sigma(X_1).
$$ 

\begin{figure}[h]
\includegraphics[width=7cm,height=6cm]{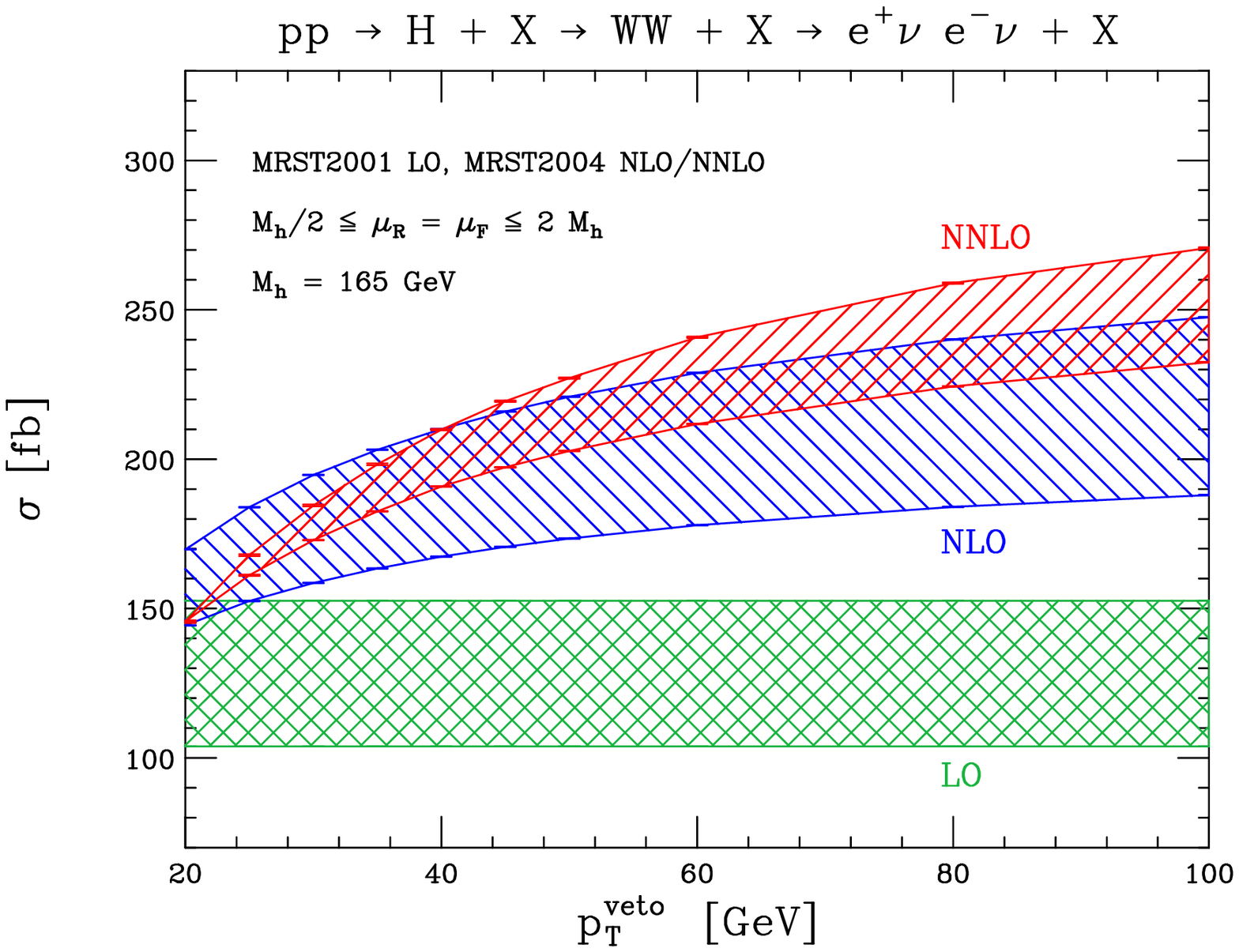}
\includegraphics[width=7cm,height=6cm]{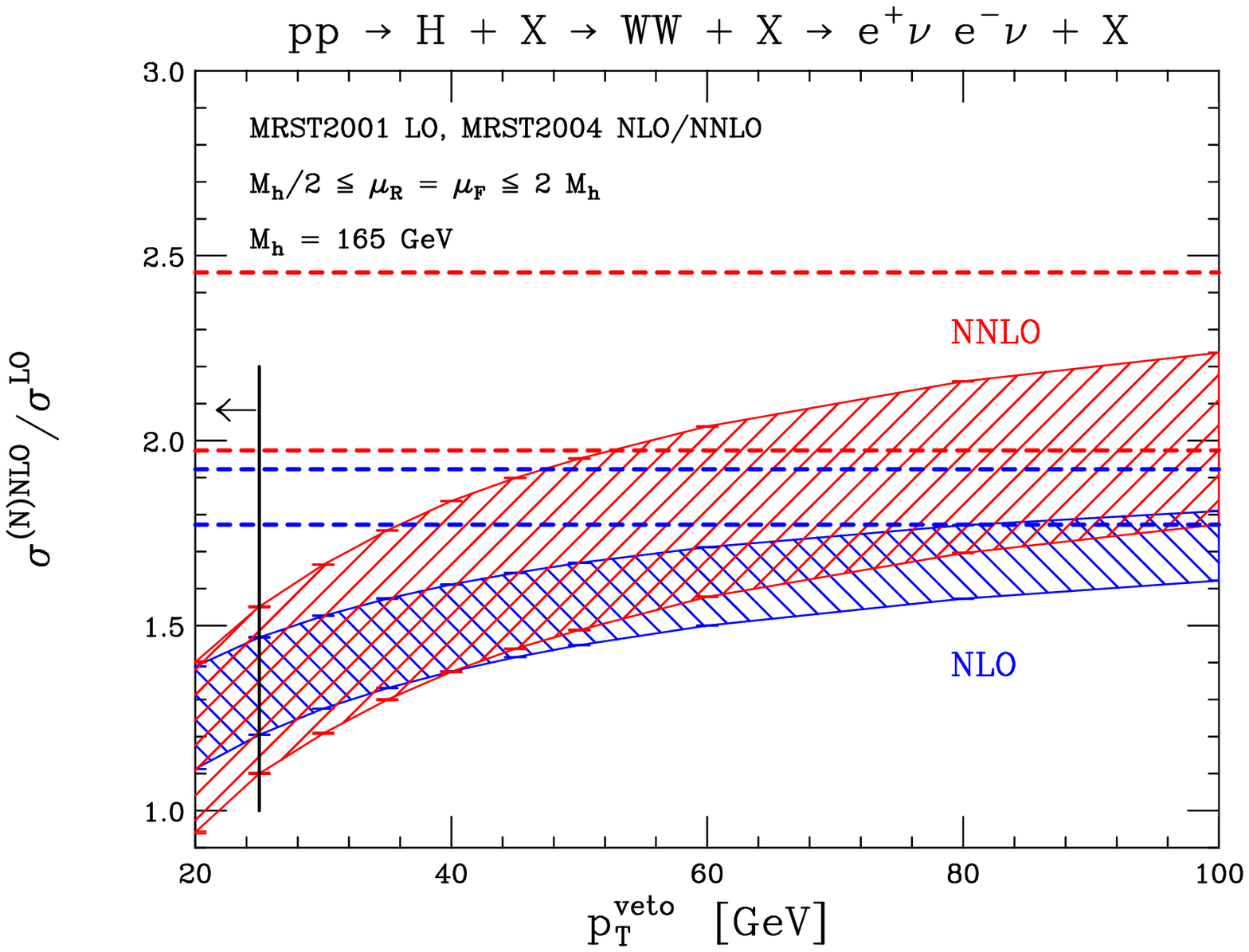}
\caption{On the left plot, the cross-section to produce a Higgs boson 
vetoing events with jets in the central region $|\eta| < 2.5$ and 
$\ptjet > \ptveto$ (no other 
cut is applied). 
On the right plot, the $K$-factor as a function of $\ptveto$. 
The dashed horizontal lines correspond to the NLO and NNLO 
$K$-factors for the inclusive cross-section. The vertical solid line 
denotes the value of $\ptveto$ in the {\it signal cuts} 
of Section~\ref{sec:selection}.}
\label{fig:jetveto}
\end{figure}

In Fig.~\ref{fig:jetveto} we re-consider the  effect of the veto on 
jets with  transverse momentum $\ptjet> \ptveto$ (see 
also ~\cite{Catani:2001cr,Anastasiou:2004xq}).  
Here, we only veto central jets with rapidity $|\eta| < 2.5$, 
while all events with jets at larger rapidity are accepted.  
Jets are defined using a cone algorithm~\cite{conealg} 
with a cone size $R= 0.4$.  
We observe that the relative magnitude of the 
NLO and NNLO contributions  depends strongly 
on \ptveto.  The NNLO cross-section increases more rapidly than the 
NLO by relaxing the veto. Fig.~\ref{fig:jetveto} demonstrates that  
the large NLO and NNLO corrections must be attributed 
to contributions from jets with large rather than small transverse momentum. 

In order to reduce the $\pp\pp \to \mathrm{t}\bar{\mathrm{t}}$ background, 
it is required to choose  a small value  of \ptveto. 
As we decrease  the value of the allowed jet transverse energy, 
the scale uncertainty at NNLO decreases. At around $\ptveto = 20\,\GeV$ the 
difference of the cross-section at $\mu = 2 \MH$ and $\mu = \frac{\MH}{2}$ 
changes sign.  In this kinematic region logarithmic 
contributions $\log(\ptveto)$  
from soft radiation  beyond NNLO
should also be examined~\cite{Catani:2001cr}. 
However, the small scale uncertainty at NNLO 
and the small magnitude of the corrections 
suggest that such logarithms have a mild effect.

\begin{figure}[h]
\begin{center}
\includegraphics[width=7cm,height=6cm]{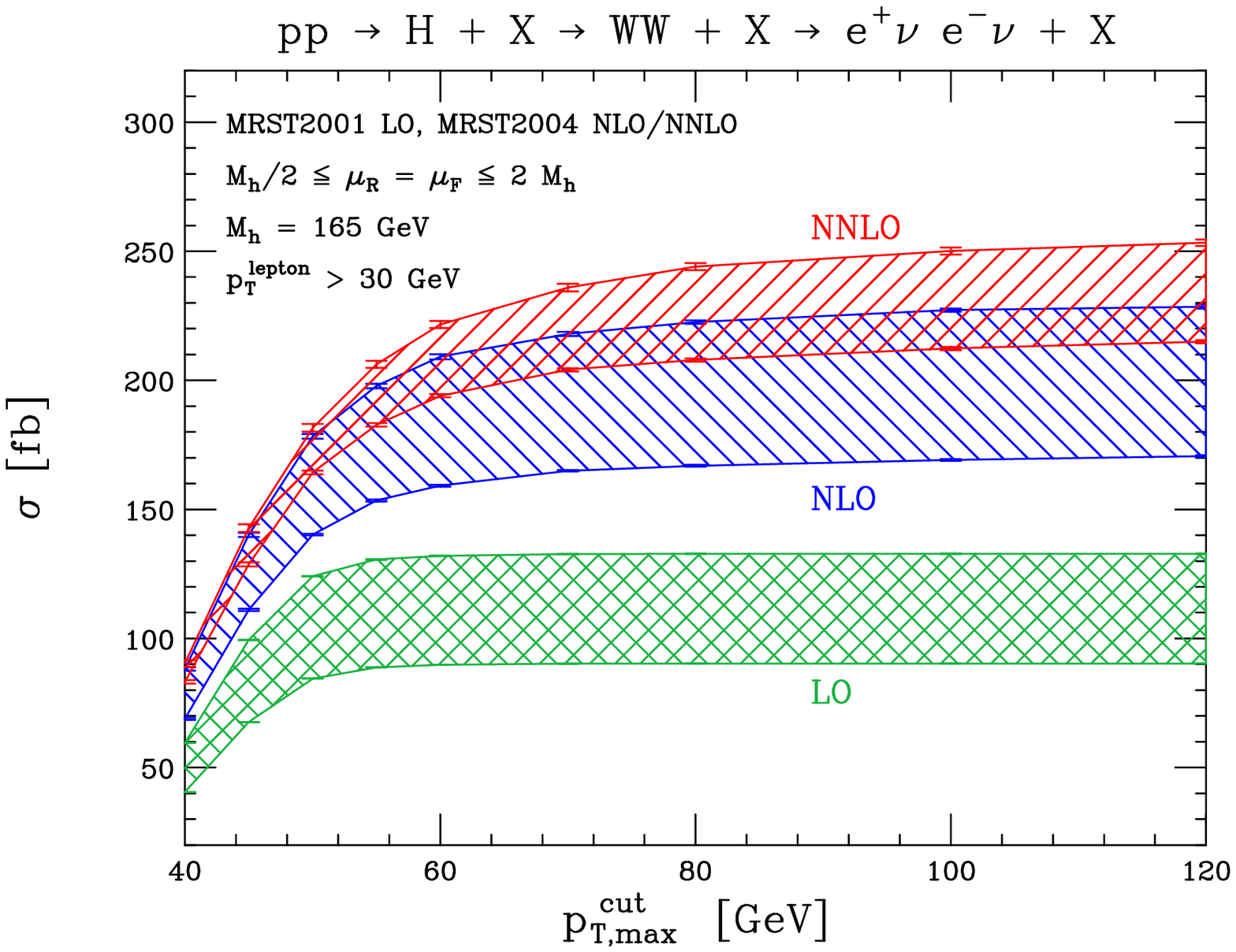}
\includegraphics[width=7cm,height=6cm]{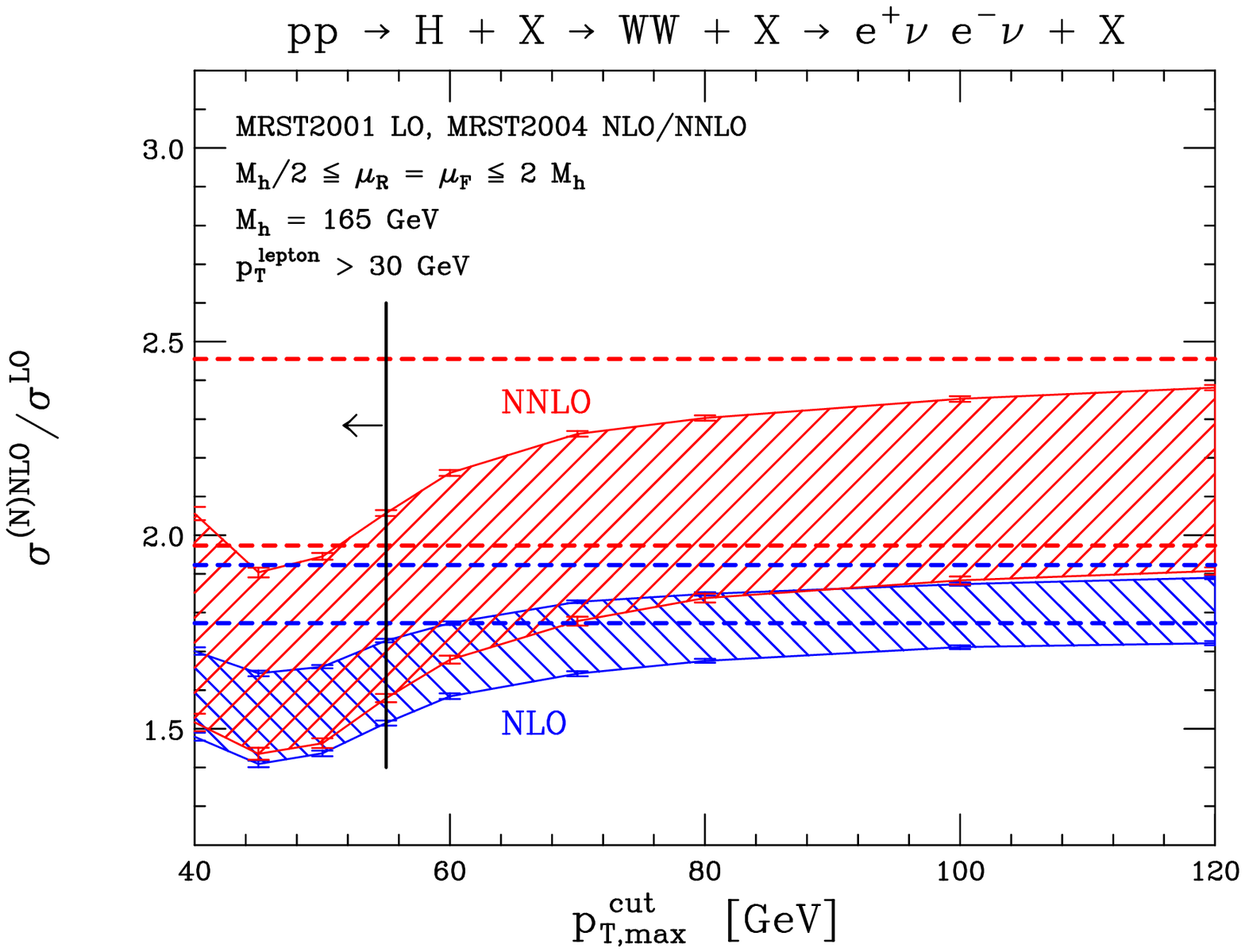}
\end{center}
\caption{
On the left plot, the cross-section  
for events where the hardest visible lepton has transverse momentum
$30 \, \GeV <\ptlep < \ptlmax$. 
On the right plot, the $K$-factor as a function of $\ptlmax$ (no other 
cut is applied). 
The dashed horizontal lines correspond to the NLO and NNLO 
$K$-factors for the inclusive cross-section. The vertical solid line 
denotes the value of $\ptlmax$ in the {\it signal cuts} 
of Section~\ref{sec:selection}.
}
\label{fig:ptmax}
\end{figure}

In Fig.~\ref{fig:ptmax} we show the cross-section after the requirement that 
the transverse momentum of the hardest visible lepton is restricted to the 
interval  $ 30\,\GeV < \ptlep < \ptlmax$.  
In Ref.~\cite{DavatzCMSnote} the upper
boundary of the allowed region was chosen as $\ptlmax = 55\,\GeV$.
At LO, only $\sim 1 \%$ of the hardest visible leptons have  
transverse momentum of $\ptlep > 55\,\GeV$.  
However, at NLO (NNLO) about $\sim 13\, (19)\, \%$ of the events lie 
above this cut. 
Thus the choice  $\ptlmax = 55\,\GeV$ removes regions of the phase-space 
that are only populated at NLO and NNLO. 
We observe that the NLO  and NNLO  $K$-factors
are smaller below this cut.   In addition, 
the scale uncertainty  drops below $12\%$ at NNLO, while the corresponding 
scale uncertainty for the inclusive cross-section is $17\%$.

\begin{figure}[h]
\begin{center}
\includegraphics[width=7cm,height=6cm]{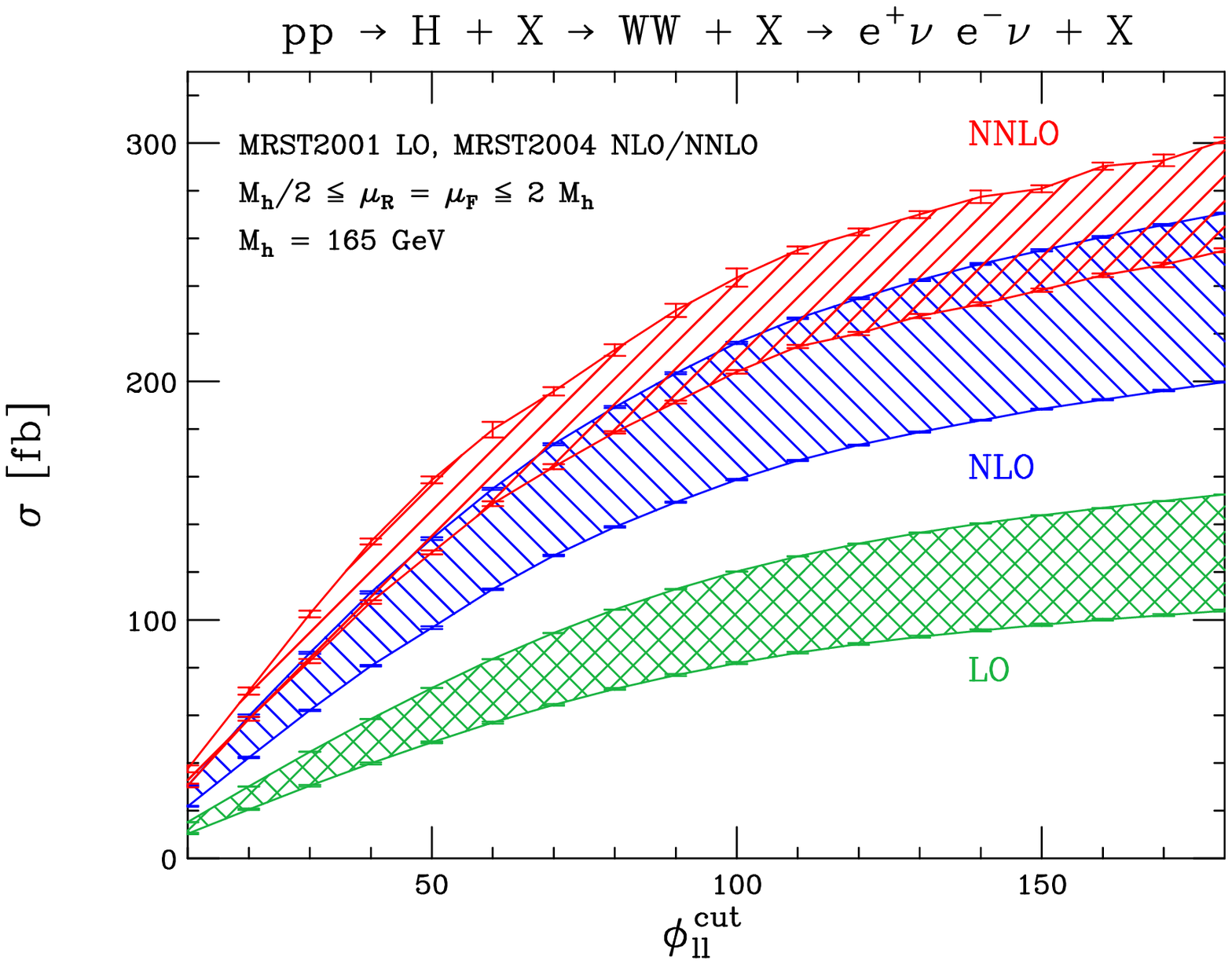}
\includegraphics[width=7cm,height=6cm]{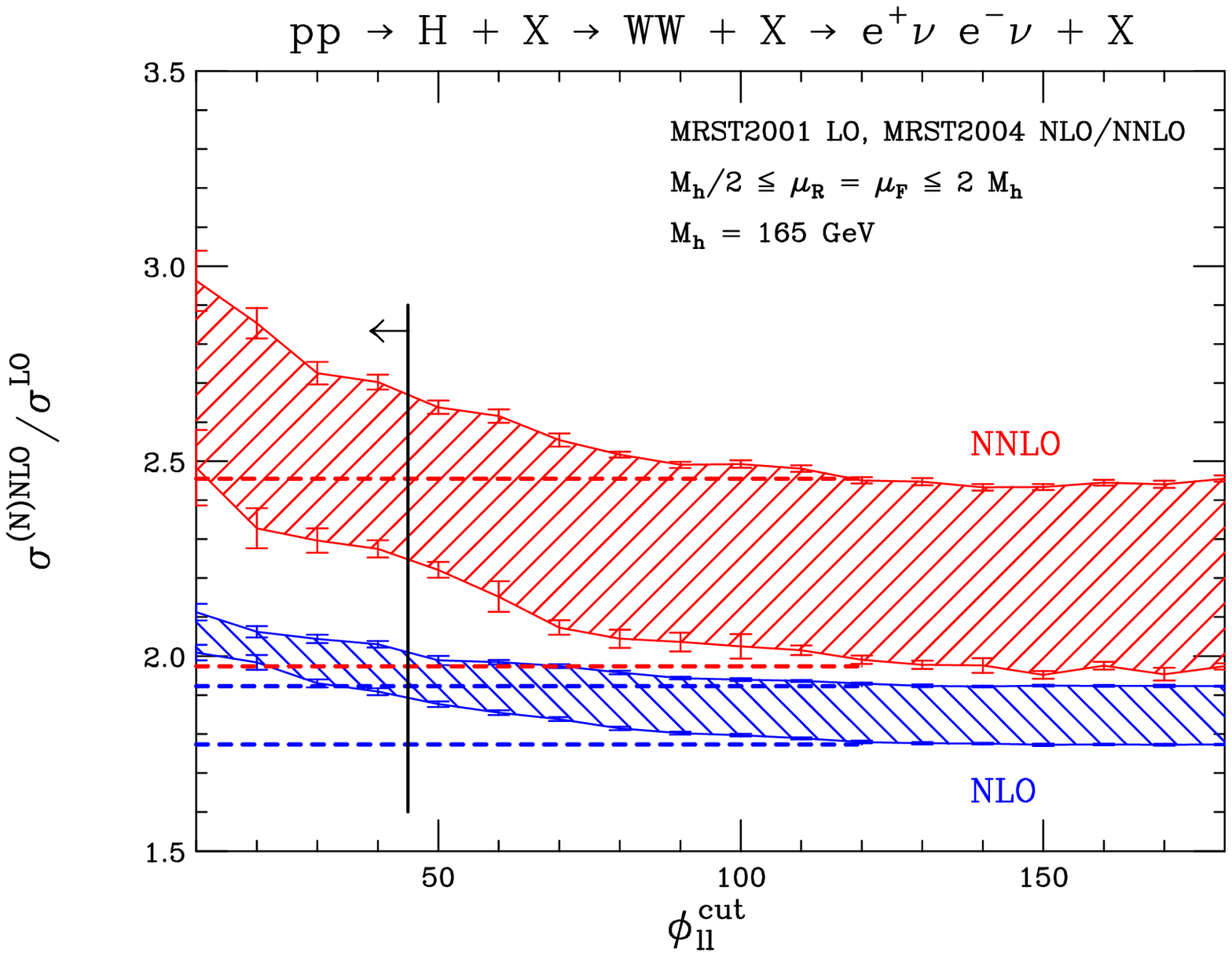}
\end{center}
\caption{
On the left plot, the cross-section  
for visible leptons with an angle on the transverse plane 
$\phi_{\lp \lp} < \phicut$. 
On the right plot, the $K$-factor as a function of $\phicut$ 
(no other cut is applied). 
The dashed horizontal lines correspond to the NLO and NNLO 
$K$-factors for the inclusive cross-section. The vertical solid line 
denotes the value of $\phicut$ in the {\it signal cuts} 
of Section~\ref{sec:selection}.
}
\label{fig:phi}
\end{figure}

A powerful discriminating variable between the signal and the $\pp\pp \to \wbo\wbo$ 
background is the opening angle $\phi_{\lp\lp}$
between  the two visible leptons in the plane transverse to the beam axis.  
In Fig.~\ref{fig:phi} we plot the cross-section for events 
with  $\phi_{\lp\lp} < \phicut$~\footnote{
We note that the distribution of the
opening angle at NNLO, using the code of~\cite{Catani:2007vq}, has been 
presented at the Les Houches workshop in June 2007~\cite{GrazziniLesHouches}. 
Qualitatively our results are similar.}. 
We observe that the NLO and especially the 
NNLO corrections are significantly larger for small angles  $\phi_{\lp\lp}$. 
For $\phicut =40^\circ$ the NNLO $K$-factor is $\sim 2.27\, (2.70)$ 
for $\mu = \frac{\MH}{2}\, (2 \MH)$. The corresponding 
$K$-factor for the inclusive cross-section is $\sim 1.97\, (2.45)$. 
The NNLO scale uncertainty for $\phicut = 40 ^{\circ}$ is  
$18.5\%$, while for the inclusive cross-section it is $\sim 17\%$. 
Thus the envisaged cut at $\phicut \sim 45^\circ$ enhances contributions 
with large perturbative corrections.  
   
\begin{figure}[h]
\begin{center}
\includegraphics[width=7cm,height=6cm]{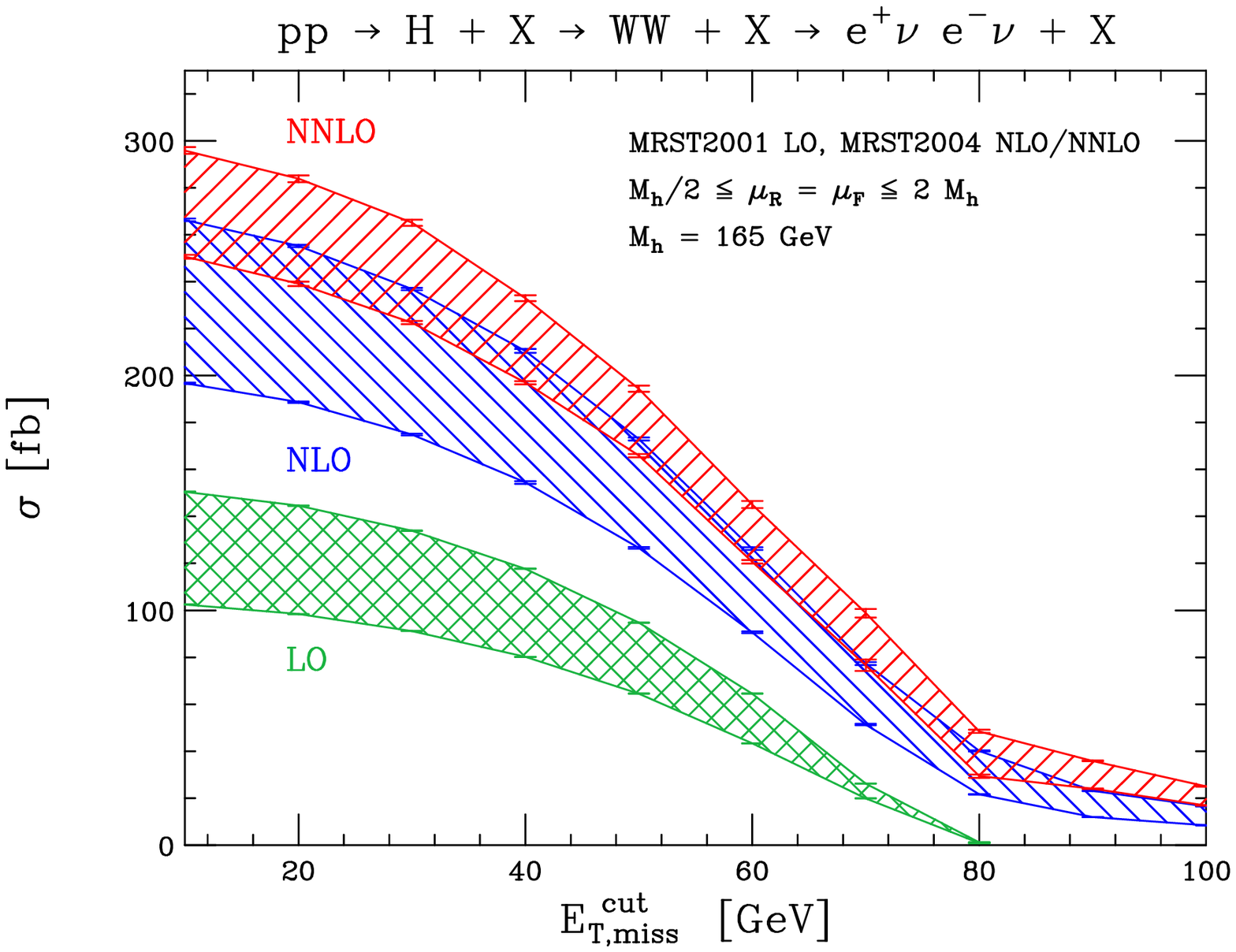}
\includegraphics[width=7cm,height=6cm]{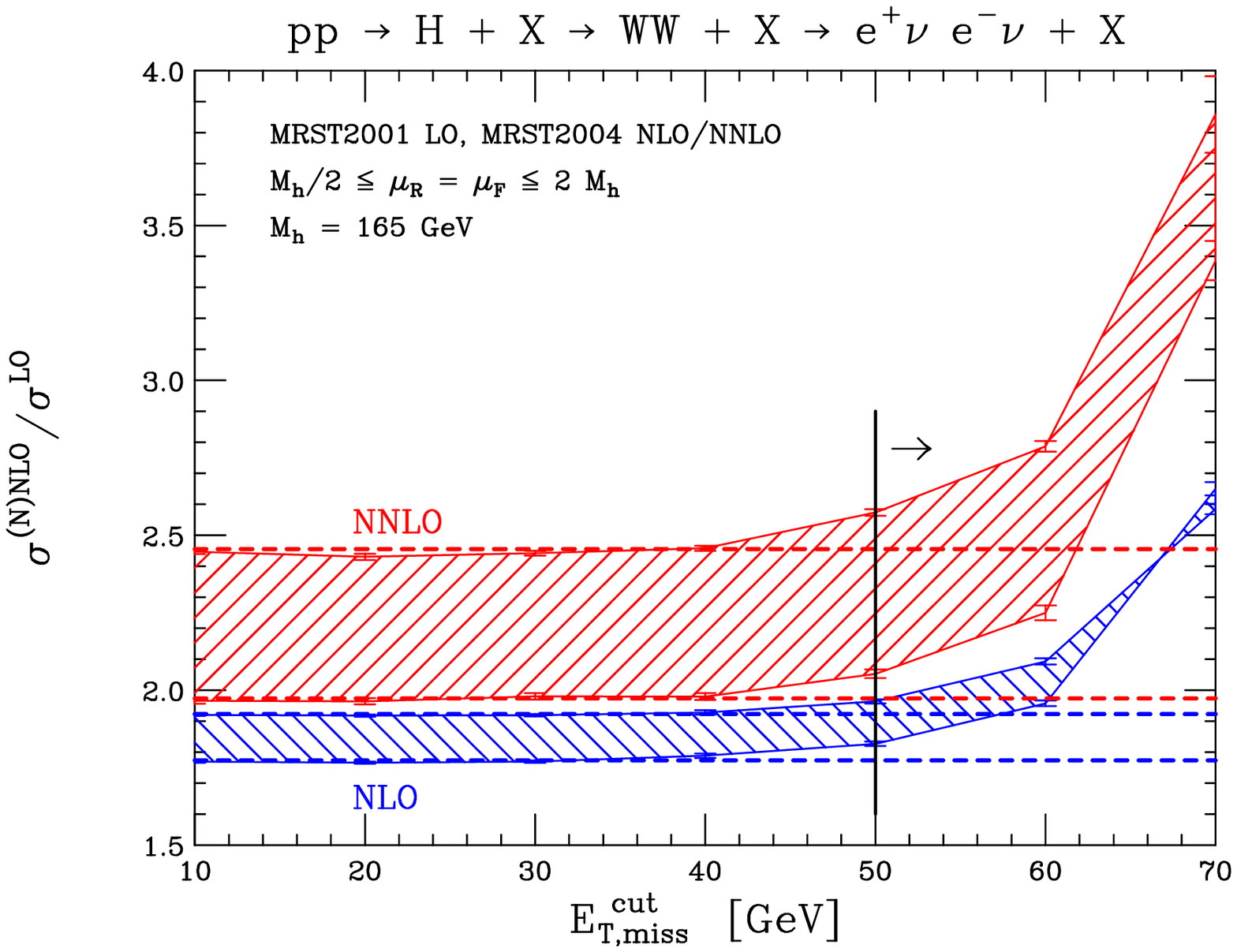}
\end{center}
\caption{
On the left plot, the cross-section  
for events with missing transverse energy $\Etmiss > \Etcut$, where 
$\Etmiss$ is computed as the transverse momentum of the neutrino pair. 
On the right plot, the $K$-factor as a function of $\Etcut$ 
(no other cut is applied). 
The dashed horizontal lines correspond to the NLO and NNLO 
$K$-factors for the inclusive cross-section. The vertical solid line 
denotes the value of $\Etcut$ in the {\it signal cuts} 
of Section~\ref{sec:selection}.
}
\label{fig:etmiss}
\end{figure}

The decay of the \wbo\ bosons produces  large missing transverse energy, \Etmiss. 
In Fig~\ref{fig:etmiss} we plot the cross-section for $\Etmiss > \Etcut$. 
 At leading order, there are no contributions 
from $\Etmiss > \Mw$. This region of the phase-space requires that the 
Higgs system is boosted with additional radiation at NLO and NNLO. 
The contribution from $\Etmiss > 80\,\GeV$, for $\mu=\frac{\MH}{2}$, 
amounts to  $0.7\%$ at LO, $\sim 14\%$  at NLO and $\sim 16\%$ at NNLO. 
The scale variation for this region of the phase-space 
is $60\%$ at NLO (essentially LO) and $49\%$ at NNLO (essentially NLO). 
By requiring very large missing transverse energy, we enhance the significance of the 
above phase-space region; the  $K$-factors  tend to increase with respect to 
the inclusive cross-section.

\begin{figure}[h]
\begin{center}
\includegraphics[width=7cm,height=6cm]{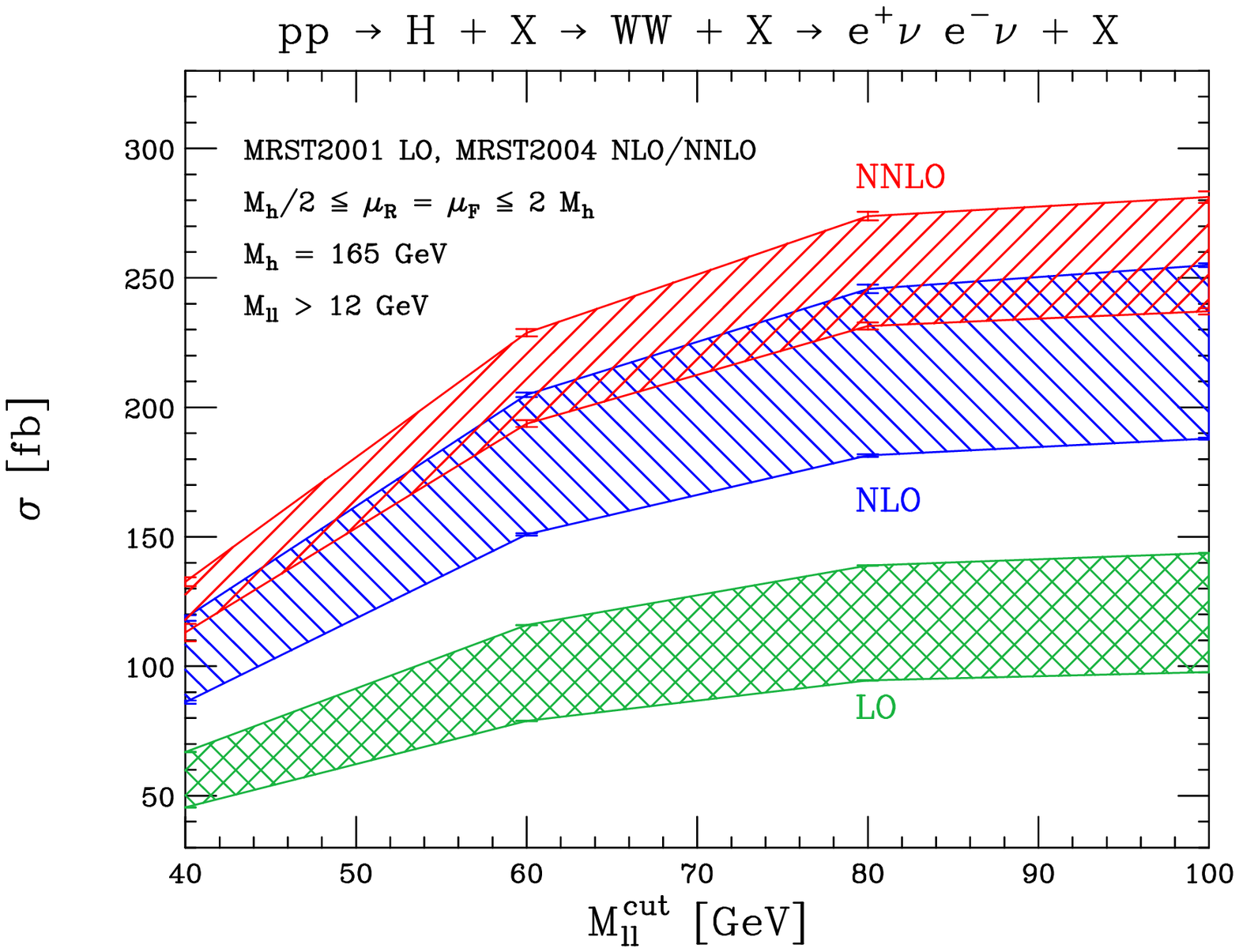}
\includegraphics[width=7cm,height=6cm]{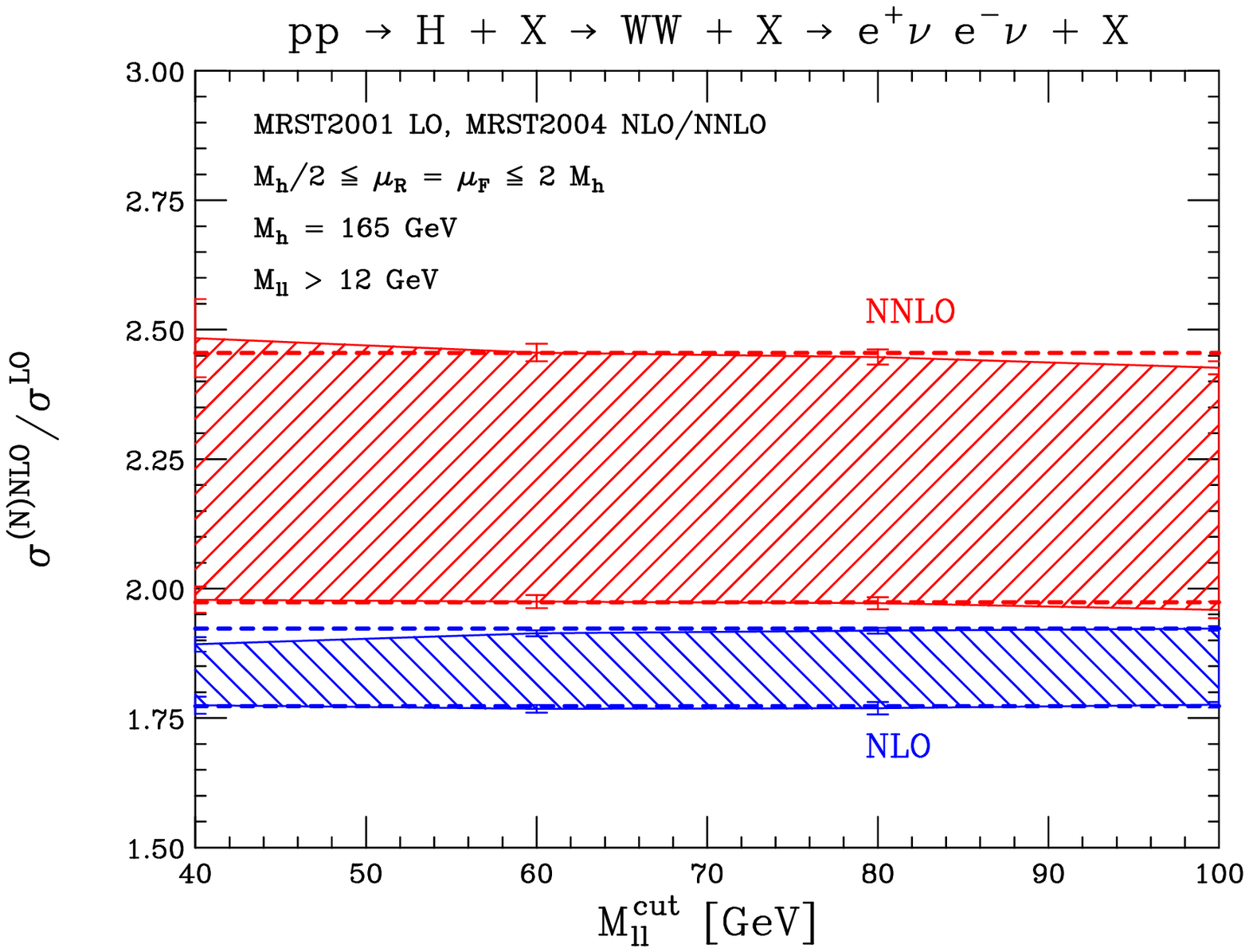}
\end{center}
\caption{
On the left plot, the cross-section  
for events with visible lepton invariant mass  $12 \, \GeV < 
M_{\lp \lp} < M_{\lp \lp}^{\mathrm{cut}}$. 
On the right plot, the $K$-factor as a function of $M_{\lp \lp}^{\mathrm{cut}}$ 
(no other cut is applied). 
The dashed horizontal lines correspond to the NLO and NNLO 
$K$-factors for the inclusive cross-section.
}
\label{fig:invmass}
\end{figure}

In Fig.~\ref{fig:invmass} we plot the cross-section for events 
with a  lepton invariant mass in  the interval 
$12\,\GeV < M_{\lp\lp} < M_{\lp\lp}^\mathrm{cut}$.  
We notice that the cross-section has a perturbative 
convergence with $K$-factors  and scale variation 
very similar to the ones for the inclusive  cross-section 
for all choices of $M_{\lp\lp}^\mathrm{cut}$. 

We have now studied  the kinematic behavior of the cross-section 
through NNLO for all variables which are subject to significant 
experimental cuts in order to optimize the signal to background ratio.  
A geometrical cut on isolating the leptons from  hadrons 
(partons in our case) rejects very few events ($\sim 1-2 \%$).

We have found that the cuts discussed above  
can change individually the $K$-factors 
and the scale variation of the cross-section.  
In the next Section we will compute the cross-section after 
applying  all the cuts  which are described in Section~\ref{sec:selection}.

\section{Signal cross-section at the  \LHC\ }
\label{sec:LHCresults}

We present now the main results of our paper, which are the cross-sections 
for the experimental cuts and parameters of Section~\ref{sec:selection}.

In Table~\ref{tab:preselection} we show the cross-section 
for the {\it pre-selection} cuts, which do not impose a jet-veto, 
for three choices  of $\mu_\mathrm{R} = \mu_\mathrm{F} = \mu$:\\ 
\begin{table}[h]
\begin{center}
\begin{tabular}{||c|c|c|c||}
\hline
$\sigma(\mathrm{fb})$         & LO     & NLO    &   NNLO  \\ 
\hline
$ \mu=\frac{\MH}{2}$ & $71.63 \pm 0.07$  
                   & $126.95 \pm 0.13$ 
                   & $140.73 \pm 0.45$
                   \\
\hline
$\mu= \MH$         & $59.40 \pm 0.06$  
                   & $108.42 \pm 0.15$
                   & $130.01 \pm 0.36$ 
                   \\
\hline
$\mu= 2 \MH$       & $49.56  \pm 0.05$  
                   & $94.33  \pm 0.13$
                   & $119.28 \pm 0.26$ 
                   \\
\hline
\end{tabular}
\end{center}
\caption{
\label{tab:preselection}
Cross-section through NNLO for the
{\it pre-selection cuts} 
of Section~\ref{sec:selection}.
}
\end{table}
The scale variation is $\sim 37\%$ at LO, $\sim 30\%$ at NLO, and
$\sim 17 \%$ at NNLO. This is a similar scale-variation 
as for the inclusive cross-section in Table~\ref{tab:inclusive}.  
The $K$-factors for the accepted cross-section are also very similar to 
the $K$-factors for the inclusive cross-section.  The 
{\it pre-selection cuts}  affect only mildly the perturbative 
convergence of the cross-section.

We find a very different behavior when the {\it signal cuts} are applied 
(Table~\ref{tab:davatzcuts}).
\begin{table}[h]
\begin{center}
\begin{tabular}{||c|c|c|c||}
\hline
$\sigma(\mathrm{fb})$         & LO     & NLO    &   NNLO  \\ 
\hline
$ \mu=\frac{\MH}{2}$ & $21.002 \pm 0.021$  
                   & $22.47 \pm 0.11$ 
                   & $18.45 \pm 0.54$
                   \\
\hline
$\mu= \MH$         & $17.413 \pm 0.017$  
                   & $21.07 \pm 0.11$
                   & $18.75 \pm 0.37$ 
                   \\
\hline
$\mu= 2 \MH$       & $14.529 \pm 0.014$  
                   & $19.50 \pm 0.10$
                   & $19.01 \pm 0.27$ 
                   \\
\hline
\end{tabular}
\end{center}
\caption{
\label{tab:davatzcuts}
Cross-section through NNLO for the 
{\it signal cuts} of Section~\ref{sec:selection}.}
\end{table}
We observe that the NLO and NNLO $K$-factors are  
small in  comparison to the corresponding $K$-factors for the 
inclusive cross-section. The relative magnitude of the NLO and NNLO 
corrections with respect to LO is similar to the 
observed $K$-factors in  Fig.~\ref{fig:jetveto} for a jet-veto value around 
$\sim 20 \, \GeV$. In addition, the scale variation is also small at 
NNLO (of similar magintude as the statistical error of our numerical integration); 
this is again similar to the pattern observed  
in Fig.~\ref{fig:jetveto} for small values of the jet-veto.

\begin{figure}[h]
\begin{center}
\includegraphics[width=12cm]{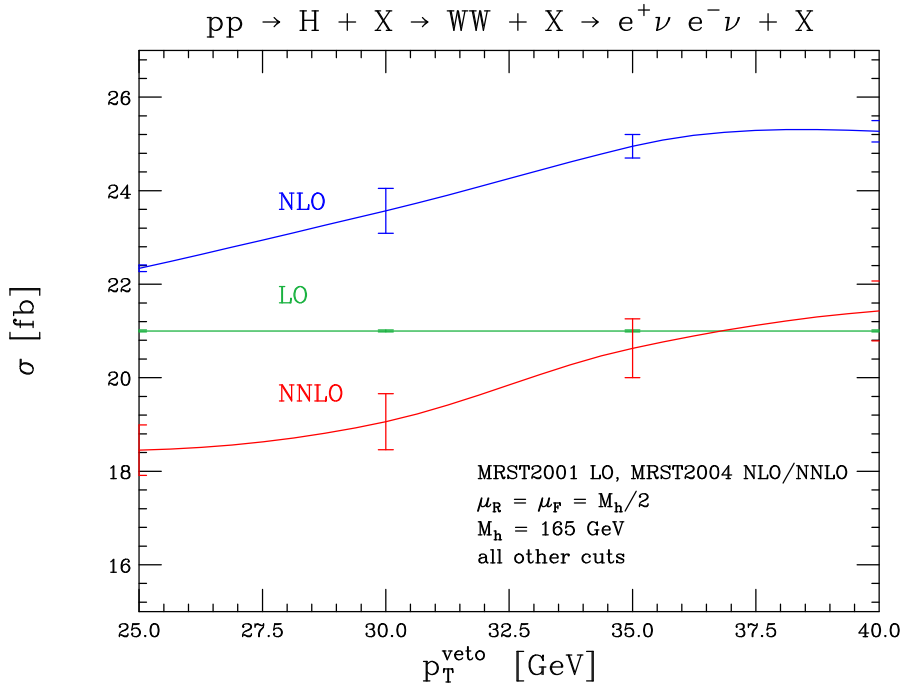}
\end{center}
\caption{The cross-section  for the {\it signal cuts} varying 
the value of the jet-veto. The increase in the cross-section 
by relaxing the jet-veto is slower than in Fig.~\ref{fig:jetveto}. 
Other cuts in addition to the jet-veto restrict the $\pt$ of central 
jets to small values.
}
\label{fig:finalcuts}
\end{figure}

The jet-veto enhances the significance of 
soft gluon radiation and  a resummation of large logarithms
may be necessary. 
We  investigate the dependence of the cross-section 
on the jet-veto in Fig.~\ref{fig:finalcuts}, where  
we have computed the 
cross-section with all {\it signal cuts} of Section~\ref{sec:selection}
and  for different values of the jet-veto $\ptveto$. 
We find that the signal cross-section at NNLO and a jet-veto value 
$\ptveto = 40\,  \GeV$ is only $13\%$ larger  than the cross-section 
for $\ptveto = 25\, \GeV$ when $\mu_\mathrm{R}=\mu_\mathrm{F}=\frac{M_h}{2}$. 
If we do not apply any other cuts except the jet-veto, the 
corresponding increase is almost double $\sim 25\%$. Therefore,  
we conclude that both the jet-veto  and 
the other cuts  constrain  central jets 
to  low transverse momentum. 

The cross-section in Table~\ref{tab:davatzcuts} 
for the {\it signal cuts} demonstrates 
a much better perturbative behavior than the inclusive cross-section. 
However, before we conclude that we have obtained a  
very precise prediction for the signal cross-section 
we would like to investigate further the importance of resummation 
effects. 
We computed the average transverse momentum of the Higgs boson to 
be $<p_{\mathrm{T}}^{\mathrm{H}}>_{ {\rm cuts}} 
\sim 15 \,\GeV$ at NNLO for $\mu_\mathrm{F} = \mu_\mathrm{R} = \frac{M_h}{2}$.  
The corresponding average for the inclusive cross-section is 
$<p_{{\mathrm T}}^{\mathrm{H}}> \sim 48\, \GeV$.   
Logarithms $\log(p_{\mathrm{T}}^{\mathrm{H}})$ 
could therefore have a larger  impact on  the  accepted 
cross-section with the {\it signal cuts} 
than the inclusive cross-section.

The existence of large logarithmic corrections 
is  not  manifest by varying  the renormalization and factorization 
scales as shown in Table~\ref{tab:davatzcuts}. 
To investigate this aspect thoroughly, we compute in Table~\ref{tab:scales}
the cross-section with the {\it signal cuts} of Section~\ref{sec:selection} 
for independent values of $\mu_\mathrm{R}$ and $\mu_\mathrm{F}$ in the interval 
$\left[ \frac{M_{h}}{4},  2 M_h \right]$.
The scale variation in this interval 
is rather small. We note that the corresponding scale variation for the 
inclusive cross-section in the smaller interval  
$\left[ \frac{M_{h}}{2},  2 M_h \right]$ is $\sim 17 \%$.  

\begin{table}[h]
\begin{center}
\begin{tabular}{|c||c|c|c|c||}
\hline
 $\sigma(\mathrm{fb})$  &$\mu_\mathrm{F} = \frac{M_h}{4}$ &$\mu_\mathrm{F} =\frac{M_h}{2}$ & $\mu_\mathrm{F} =M_h$ 
& $\mu_\mathrm{F} =2 M_h$  
\\ 
\hline
\hline
       
$\mu_\mathrm{R}  = 2M_h$ &$ 17.89 \pm 0.27 $&$ 18.27 \pm 0.29 $&$ 18.97 \pm 0.29 $&$ 19.01 \pm 0.27 $\\ 
\hline

$\mu_\mathrm{R}  = M_h$ &$ 18.68 \pm 0.90$ &$18.33 \pm 0.40$&$ 18.75 \pm 0.37$&$ 19.87 \pm 0.42$ \\
\hline

$\mu_\mathrm{R}  =  \frac{M_h}{2}$ & $18.84 \pm 0.60$ & $18.45 \pm 0.54$ & $17.52 \pm 0.93$ & $18.10 \pm 0.63$ \\
\hline

$\mu_\mathrm{R}  = \frac{M_h}{4} $ & $16.82 \pm 0.94$ & $18.40 \pm 1.00$ & $16.06 \pm 0.94$ & $15.45 \pm 0.98$ \\
\hline
\hline
\end{tabular}
\end{center}
\caption{
\label{tab:scales}
NNLO cross-section for the {\it signal cuts} and independent values of 
the renormalization scale $\mu_\mathrm{R}$ and the factorization scale $\mu_\mathrm{F}$.
}
\end{table}

We can quantify the effect of $\pt$ logarithms and the 
need for resummation comparing our NLO and NNLO 
predictions with the prediction from the 
parton-shower generator  MC@NLO~\cite{mcnlo,spincor}. 
A comparison of the accepted cross-sections 
with the cuts of Section~\ref{sec:selection} is not immediately 
possible, since the spin correlations in 
the $\HH \to  \wbo \wbo \to \lp \nu \lp \nu$ decay are not 
treated fully in HERWIG~\cite{herwig}. 
However,  a similar comparison has been made in 
~\cite{Davatz:2006ut} for the Higgs boson cross-section when only a 
jet-veto is applied at $\ptveto =30\, \GeV$. It was found that 
the MC@NLO result is  $\sim 26 \%$ smaller than the NLO. The NNLO result 
is smaller than NLO by only about $\sim 9\%$. 
If one normalizes the MC@NLO to the NNLO inclusive cross-section, 
the accepted cross-sections for  MC@NLO and NNLO after the jet-veto 
are close; it was found in~\cite{Davatz:2006ut} that the MC@NLO 
efficiency is $\sim 51 \%$, while the NNLO efficiency is $\sim 54 \%$. 
We note that the effect of resummation in comparision to NLO calculations 
for $\pp\pp \to \HH \to \wbo\wbo$ has been studied in~\cite{cao}, 
however the cuts applied there did not include a jet-veto.

Our NNLO result, which is very close to NLO, exhibits a remarkable 
stability with  varying the renormalization and 
factorization scales; this alludes, without proving it, to small numerical 
coefficients of logarithmic terms.  In addition, in the presence of the 
jet-veto only, the MC@NLO and NNLO efficiencies are not very different
suggesting that the NNLO result has captured to a large extend 
the effect of low $\pt$ radiation. In a hypothetical ``MC@NNLO'' calculation 
the difference to our NNLO result could be even  smaller. 
However, in order to verify this intuition,  
a better understanding of resummation effects in the presence of 
all experimental cuts is indispensable. 

It is interesting to investigate whether a ``loosening'' of the 
experimental cuts could alter the perturbative behavior of the 
cross-section.
Changes in the experimental cuts influence the background cross-sections 
more significantly than the signal cross-section. 
Given the complexity of the combined background $\pp\pp\to \mathrm{t} \bar{\mathrm{t}}$ 
and $\pp\pp \to \wbo \wbo$ processes, it appears to us that 
there is little freedom for major changes without spoiling the 
estimated $S/B$ ratio in ~\cite{Davatz:2006ja}.  
We apply the following changes to the {\it signal cuts} of 
Section~\ref{sec:selection}: 
\begin{itemize}
\item apply a less restrictive jet-veto $\ptveto = 35 \, \GeV$;
\item require smaller $\Etmiss > 45 \, \GeV$;
\item allow a larger lepton invariant mass 
$12\,\GeV < M_{\lp\lp} < 45\,\GeV$;
\item allow larger lepton angles $\phi_{\lp\lp} < 60^\circ$; 
\item do not restrict the upper value of the $\pt$ of  the 
hardest lepton, 
$\ptlep > 30 \, \GeV$.
\end{itemize}
For these new cuts the average momentum of the Higgs boson is 
only by little larger, $<p_\mathrm{T}^{\mathrm{H}}> \sim 18 \, \GeV$. 
We find the new cross-section in Table~\ref{tab:newcuts}.
\begin{table}[h]
\begin{center}
\begin{tabular}{||c|c|c|c||}
\hline
$\sigma(\mathrm{fb})$         & LO     & NLO    &   NNLO  \\ 
\hline
$ \mu=\frac{\MH}{2}$ &   $ 28.811 \pm 0.028$  
                     &   $ 35.81  \pm 0.22 $ 
                     &   $ 32.48  \pm 0.52 $
                   \\
\hline
$\mu= \MH$         & $ 23.884 \pm 0.023$  
                   & $ 32.53  \pm 0.16 $
                   & $ 31.59  \pm 0.38 $ 
                   \\
\hline
$\mu= 2 \MH$         & $ 19.933  \pm  0.019 $  
                     & $ 29.53  \pm   0.15  $
                     & $ 31.45  \pm   0.26  $ 
                   \\
\hline
\end{tabular}
\end{center}
\caption{
\label{tab:newcuts}
Cross-section through NNLO for {\it loose signal cuts}.}
\end{table} 
We find once again very small NNLO corrections  
with respect to the NLO cross-section.
The scale variation is very small and 
remains comparable to our Monte-Carlo integration error.

The NNLO $K$-factor for the cross-section with the {\it signal cuts} of 
Table~\ref{tab:davatzcuts} is $~ 0.9 - 1.3$ depending on the scale choices. 
One must be careful if this $K$-factor is applied to rescale the result 
of a leading order parton-shower generator.  At LO in fixed order 
perturbation theory, all events have Higgs $\pt =0$; a jet-veto 
has a $100\%$ efficiency. Parton-shower event generators produce 
an extended $\pt$ spectrum, and have a significantly smaller 
efficiency; for example, the efficiency of Pythia~\cite{pythia} 
with a jet-veto at 
$\ptveto = 30 \, \GeV$ is about $50\%$~\cite{Davatz:2006ut}. 
The appropriate factor for re-weighting  LO event generators is: 
$$
K_{\mathrm{NNLO}} \times \frac{\mathrm{efficiency(LO)}}{\mathrm{efficiency(MC)}}
$$
This factor yields qualitatively similar results as in 
Refs~\cite{Davatz:2006ut,Davatz:2004zg}. However, we have not yet made 
a consistent comparison of our NNLO result for the 
signal cross-section and existing predictions 
from studies based on re-weighting~\cite{Davatz:2006ut,Davatz:2004zg}.

\section{Conclusions}
\label{sec:TheEnd}

We have performed a first calculation of kinematic distributions 
and the cross-section with experimental cuts in NNLO QCD for the process {{\ppHWWlept}}. 
For this purpose, we  have extended the Monte-Carlo program {{\FEHiP}}~\cite{fehip}, 
by including the matrix-elements for the decay of the Higgs boson
and parallelizing the evaluation of 
sectors~\cite{Anastasiou:2003gr}. 

We have observed that many kinematic distributions exhibit $K$-factors and 
scale variations which are qualitatively different than in the inclusive 
cross-section.  As a consequence, only when mild ({\it pre-selection}) 
cuts are applied the cross-section receives large perturbative 
corrections through NNLO  as for the inclusive 
cross-section. 
In contrast, for the selection cuts which are designed to isolate the 
Higgs boson signal from the background, we find small NNLO corrections 
and a very good stability  with varying the renormalization and 
factorization scales. 

The experimental cuts restrict the phase-space to events with small 
transverse momentum for the Higgs boson. The effect of resummation should be 
investigated thoroughly in future works. However, 
large logarithms do not  
become manifest when varying the renormalization and factorization 
scales, and the efficiencies at NNLO and MC@NLO for a typical jet-veto 
cut  differ by less than $6 \%$.  

We find that the NNLO $K$-factors for the signal cross-section after 
the application of selection cuts are very different than the $K$-factor 
for the inclusive cross-section.
When the NNLO $K$-factors, which we have computed 
here, are used to re-weight leading order event generators, 
the large ratio between the efficiencies of the fixed order 
LO result and the prediction of the generators should also be taken 
into account.  

\section*{Acknowledgements}

We are grateful to Alejandro Daleo, Michael Dittmar and 
Giulia Zanderighi for discussions and very 
important observations. 
We thank Bryan Webber for his comments and 
communications about the MC@NLO and HERWIG event generators.  
We thank Giovanna Davatz for communicating to us research in  
her Ph.D thesis and collaboration on an earlier related project. 
We are grateful to Thomas Gehrmann, Michele della Morte, 
Filip Moortgat and Thomas Punz  
for their help in securing adequate computing resources and useful 
discussions.   
CA is grateful to Kirill Melnikov and Frank Petriello for a fruitful 
collaboration in writing the \FEHiP\ program. 
We thank the groups of theoretical physics 
at the University of Z\"urich and  
at ETH Z\"urich for providing computing resources to us. 
This research was supported in part by the Swiss National Science Foundation
(SNF) under the contract 200020-113378/1.


\bibliographystyle{JHEP}
 
\end{document}